\documentclass{emulateapj}

\newcommand{\Msun}{\ensuremath{M_\odot}}

\shorttitle{X-ray Irradiated AGN Disks} 
\shortauthors{Chang, Quataert \& Murray}

\begin{document}

\title{From Thin to Thick: The Impact of X-ray Irradiation on
Accretion Disks in AGN} 

\author{Philip Chang\altaffilmark{1,2}, Eliot Quataert\altaffilmark{1}, and Norman Murray\altaffilmark{3,4}}
\altaffiltext{1} {Astronomy Department and Theoretical Astrophysics
  Center, 601 Campbell Hall, University of California, Berkeley, CA
  94720; pchang@astro.berkeley.edu, eliot@astro.berkeley.edu}
\altaffiltext{2} {Miller Institute for Basic Research}
\altaffiltext{3} {Canadian Institute for Theoretical Astrophysics, 60
  St. George Street, University of Toronto, Toronto, ON M5S 3H8,
  Canada; murray@cita.utoronto.ca. }
\altaffiltext{4} {Canada Research Chair in Astrophysics}

\begin{abstract}
  We argue that the X-ray and UV flux illuminating the parsec-scale
  accretion disk around luminous active galactic nuclei (AGN) is
  super-Eddington with respect to the local far-infrared dust opacity.
  The far infrared opacity may be larger than in the interstellar
  medium of the Milky Way due to a combination of supersolar
  metallicity and the growth of dust grains in the dense accretion
  disk. Because of the irradiating flux, the outer accretion disk
  puffs up with a vertical thickness $h\sim R$.  This provides a
  mechanism for generating a geometrically thick obscuring region from
  an intrinsically thin disk. We find obscuring columns $\sim 10^{22}
  - 10^{23}\,{\rm cm}^{-2}$, in reasonable agreement with
  observations.
\end{abstract}


\section{Introduction}

The Seyfert unification model postulates the existence of a dusty
geometrically thick obscuring region (the ``torus'') in order to
account for many of the observed differences among various classes of
AGN (Antonucci 1993).  The large infrared fluxes from quasars also
provide evidence for obscuration of the central source, with covering
fractions of 10-50\% (Sanders et al. 1989).  Recent observations with
SDSS have directly detected a large number of Type 2 (obscured)
quasars (e.g. Zakamska et al. 2003, 2005; Ptak et al. 2006).

The origin of the geometrically thick obscuring material in AGN
remains uncertain.  The host galaxy may provide the obscuration as in
the case of MCG-6-30-15 (Ballantyne, Weingartner, \& Murray 2003).  A
warped disk is directly implicated in the low-luminosity AGN NGC 4258
(Greenhill et al. 1995) and may be relevant more generally (Sanders et
al. 1989).  Alternatively, the obscuring material may arise in an
outflow from the underlying accretion disk (e.g., Konigl \& Kartje
1994).  Energy injection by star formation (e.g., Wada \& Norman 2002;
Thompson et al. 2005) or viscous stirring in a clumpy accretion disk
(Krolik \& Begelman 1988) may also help generate the large random
motions necessary to maintain $h \sim r$.  Lastly, Pier \& Krolik
(1992) pointed out that, for luminous AGN, radiation pressure from the
central AGN could help maintain the vertical thickness of obscuring
material.

In this paper, we present a model for how geometrically thick
obscuring material might arise from X-ray illumination incident on an
initially thin AGN disk (\S~\ref{sec:illuminated disk}).  This model
is complementary to that of Pier \& Krolik (1992); we focus on how
X-ray irradiation can thicken an underlying thin disk, while they
highlighted the importance of radiation pressure support for already
geometrically thick material.
In \S~\ref{sec:dust}, we argue that the far infrared opacity in AGN
disks is larger than that in the interstellar medium (ISM) of the
Milky Way, which increases the dynamical importance of X-ray and UV
irradiation.  The dust opacity is enhanced due to 1.  supersolar
metallicity and 2. larger dust grains. We discuss the evolution of
dust grains in AGN disks via accretion of metals from the gas and dust
grain coagulation.  We show that the conditions in a thin AGN disk
appear conducive to grain growth (as also suggested by, e.g., Laor \&
Draine 1993 and Maiolino et al.  2001ab).
Finally, in \S~\ref{sec:discussion} we discuss our results and their
observational implications.

\section{X-ray Heated AGN Disks}\label{sec:illuminated disk}

\begin{figure}
\epsscale{1.0} 
\plotone{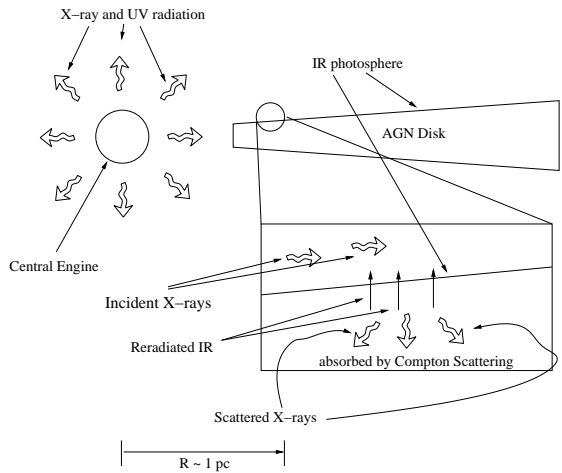}
\caption{Schematic Diagram of an X-ray Illuminated AGN Disk}
\label{fig:model}
\end{figure}

Here we consider the impact of X-ray illumination on the structure of
AGN disks.  Our physical picture is as follows (see Fig.
\ref{fig:model}).  UV and X-ray radiation from the central source are
incident at a grazing angle on a region of a thin disk which has a
thickness $h/R$.  Due to the high opacity of dust to UV photons, the
incident UV flux is absorbed well above the IR photosphere of the
disk.  UV irradiation thus tends to make the disk more isothermal and
does not provide an internal radiation pressure gradient which can
help support the disk.  X-rays in the 10-100 keV range, on the other
hand, are electron-scattered in the outer atmosphere of the disk with
approximately half of the X-rays being downscattered.  If the
underlying accretion disk is sufficiently Compton thick, most of the
downscattered X-rays lose their energy via Compton scattering below
the IR photosphere.  This energy is then re-radiated in the IR,
diffuses towards the surface of the disk, and provides a radiation
pressure gradient which is dynamically important if the energy
deposition via X-rays is sufficiently high.  We show that as a result
of this X-ray irradiation, the entire disk cannot remain thin.
Instead, the outer atmosphere puffs up to $h \sim R$ producing
significant obscuration along most lines of sight.  As discussed later
in \S 4, our calculation does not quantitatively describe the final
structure of the resulting geometrically thick atmosphere, which
requires a multi-dimensional calculation that incorporates both UV and
X-ray irradiation.

\begin{figure}
\epsscale{1.0} 
\plotone{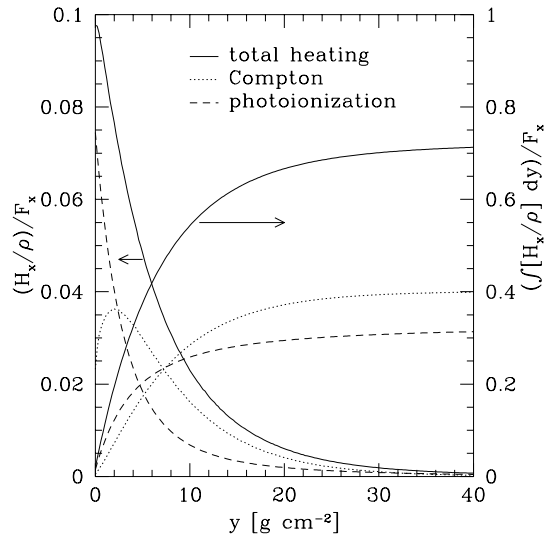}
\caption{Local heating rate ($H_x/\rho$) per unit mass and integrated
  heating rate ($\int (H_x/\rho) dy$) as a function of column $y$
  (measured downward into the disk) for a slab irradiated by a
  power-law X-ray spectrum with a photon index of $1.7$ and a high
  energy cutoff of 100 keV.  Arrows indicate the vertical axis
  associated with each solid line.  The heating rates are normalized
  to the total x-ray flux incident on the column ($F_x$) so that
  $(\int H_x dy)/F_x$ is the fraction of the incident flux that has
  been absorbed at a given depth $y$.  The total heating rate, Compton
  scattering heating, and photoionization heating are given by the
  solid-lines, dotted lines and dashed lines, respectively.}
\label{fig:xray_heating}
\end{figure}

We now calculate the critical luminosity so that X-ray irradiation
becomes dynamically important. The incident X-ray flux on the
initially thin disk is
\begin{equation}
F_X = \frac {L_X} {4\pi R^2} \frac h R,
\end{equation}
where $L_X$ is the X-ray luminosity from the central source and $R$ is
the radius of the disk.  The X-ray opacity is provided by electron
scattering and photoionization.  For X-ray energies above $\approx
10\,{\rm keV}$, electron scattering dominates.  At these energies, an
x-ray photon incident on a disk with scale height $h/R$ is scattered
at a column of $1/\kappa_{e} (h/R)$, where $\kappa_e$ is the electron
scattering opacity.  Since the scattering is roughly isotropic, half
of the X-rays are scattered out to infinity and half are scattered
down into the disk.  The half that are scattered downward heat the
disk via photoionization and Compton scattering.  We adopt the
photoionization cross section from Maloney, Hollenbach, \& Tielens
(1996) which gives $\sigma = 4.4\times 10^{-22} \left(E/{1\,{\rm
keV}}\right)^{-8/3}\, {\rm cm}^{-2}$ for $E>7$ keV, where $E$ is the
energy of the X-ray photon.  We calculate the efficiency of this
heating for an incident spectrum typical of AGN (photon index $\alpha
= 1.7$) as a function of depth and show the results in Figure
\ref{fig:xray_heating}.  The total energy absorbed by the disk is a
function of its Compton thickness, but above a total column,
$y_{\rm tot} = \int\rho dz$, of $20\,{\rm g\,cm^{-2}}$, the heating
saturates.  For these columns, $\approx 2/3$ of the energy of the
downward scattered X-rays heats the disk, with most of the energy
being deposited in the first few Thomson depths.  Models of accretion
disks in luminous AGN predict surface densities of $\sim 10-10^3$ g
cm$^{-2}$ on parsec scales (Sirko \& Goodman 2003; Thompson et al.
2005), so X-ray heating is expected to be quite important.  These
X-rays are reprocessed into the infrared where the opacity is
substantial.

Under the conditions of interest, the radiation pressure gradient
typically exceeds the gas pressure gradient near the IR photosphere.
In order to support the photosphere to a height $h$, the IR flux,
$F_{\rm IR}$, through the disk must satisfy the dust Eddington limit
($F_{\rm IR} \approx F_{\rm Edd,d}$) or
\begin{equation}\label{eq:edd}
F_{\rm IR} \approx F_{\rm Edd,d} = \frac {\Omega^2 h c}{\kappa_{d}},
\end{equation}
where $\kappa_d$ is the dust IR opacity.  Since the net flux radiated
by the IR photosphere is due to the fraction of the incident X-ray
flux deposited beneath the photosphere, we also have
\begin{equation}\label{eq:irr}
F_{\rm IR} = \epsilon\frac {L_X}{4\pi R^2} \frac h R,
\end{equation}
where $\epsilon \approx 1/3$ is the efficiency with which X-ray flux
is absorbed deep in the disk, which we estimate from the half of the
X-rays that are downward scattered and the two-thirds of this energy
that is absorbed.  For a warped disk, all of the incident X-ray may
heat the disk, not just the scattered radiation.  Equating equations
(\ref{eq:edd}) and (\ref{eq:irr}), we find a critical luminosity
\begin{equation}\label{eq:Lcrit}
  L_{\rm X,\, crit} = \epsilon^{-1}\frac {4\pi G M_{\rm BH} c} {\kappa_{d}} =
  \epsilon^{-1}L_{\rm Edd} \frac {\kappa_{e}}{\kappa_{d}},
\end{equation}
where $L_{\rm Edd} = 4\pi GM_{\rm BH}c/\kappa_e$ is the electron
scattering Eddington limit and $M_{\rm BH}$ is the mass of the central
black hole. For $L_X > L_{\rm X,\, crit}$, the disk is locally
unstable to puffing up because the energy deposited below the IR
photosphere exceeds the local dust Eddington limit.

The critical luminosity $L_{\rm X,\, crit}$ is sensitive to the
details of the dust opacity in AGN disks through
$\kappa_{e}/\kappa_{d}$.  For the conditions of interest $T\sim
300-1000$, K and normal ISM dust is estimated to have a Rosseland mean
opacity between $\kappa_d \approx 5$ (Semenov et al. 2003) and
$\kappa_d \approx 15\,{\rm cm^2\,g^{-1}}$ (Draine 2003).  In the next
section, we argue that the far infrared opacity in the near-AGN
environment may be as large as $\kappa_d \sim 50\,{\rm cm^2\,g^{-1}}$
due to increased metallicity and grain growth in AGN disks.  In this
case, $\kappa_{e}/\kappa_{d} \approx 0.01$ which in turn implies
$L_{\rm X, crit}/L_{\rm Edd} \approx 0.03$.  Bright Seyferts and
quasars have X-ray luminosities that are roughly 0.01-0.1 of the
(electron-scattering) Eddington luminosity, so a large number of
systems are near the critical luminosity needed to have X-ray heating
significantly affect the structure of a thin dusty parsec-scale disk
(\S 4).

To understand the impact of X-ray irradiation in more detail, we
calculate the vertical structure of an irradiated thin disk.  We begin
with the equation of hydrostatic balance,
\begin{equation} \label{eq:dPgdz}
\frac {dP_{\rm gas}} {dz}= -\rho \left(g - \frac{\kappa_d F} {c}\right),
\end{equation}
and radiative balance,
\begin{equation} \label{eq:dPrdz} 
\frac {dP_{\rm rad}} {dz}= - \frac{\kappa_d F \rho} {c},
\end{equation}
where $P_{\rm gas} = \rho k_B T/m_p$ is the gas pressure, $T$ is the
temperature, $P_{\rm rad} = aT^4/3$ is the radiation pressure, $F$ is
the flux, $z$ is the height defined from the midplane, $g = 2\pi G \Sigma +
\Omega^2 z$ is the local gravitational acceleration, and $\Omega =
\sqrt{GM_{\rm BH}/R^3}$ is the local Keplerian frequency.  Note we
define the disk surface density $\Sigma=\int\rho dz =0$ at the midplane
here and is defined upward.  For the Rosseland mean opacity of dust,
we assume a simple model for numerical convenience: we take a constant
value of $\kappa_d = 50$ cm$^2$ g$^{-1}$ between $T=100$ K and
$T=T_{\rm sub} = 1500$ K (see eq. [\ref{eq:opac}]).  We model the
rapid drop in opacity as the dust sublimates using $\kappa_d \propto
\exp\left(-(T-T_{\rm sub})/\Delta T\right)$ for $T>T_{\rm sub}$, where
$\Delta T \approx 100$ K defines the rough width of the transition.
None of the qualitative features of our model depend on the choice of
$\kappa_d = 50$ cm$^2$ g$^{-1}$ at low $T$, although the quantitative
importance of irradiation does depend on the absolute normalization of
$\kappa_d$ (as discussed above in the context of $L_{\rm X,crit}$).

The impact of X-ray heating on the structure of the disk can be
modeled as a source term in the equation for the flux:
\begin{equation}\label{eq:dfdz}
\frac {dF} {dz} = H(z)
\end{equation}
where $H(z)$ is the heating function shown in Figure
\ref{fig:xray_heating}.  To calculate the structure of the irradiated
disk, we integrate equations (\ref{eq:dPgdz}), (\ref{eq:dPrdz}), and
(\ref{eq:dfdz}) from $z=0$ to $z=z_0$, where $z_0$ is the IR
photosphere.  We enforce $F(z=0)=0$ at the midplane and the
photospheric condition at $z=z_0$, $\tau \approx \kappa \rho h_p =
2/3$, where $\tau$ is the optical depth taken from infinity and $h_p$
is the local scale height of the photosphere.  Numerically we specify
the height of the IR photosphere and compute the required incident
X-ray flux.

\begin{figure}
\epsscale{1.0} 
\plotone{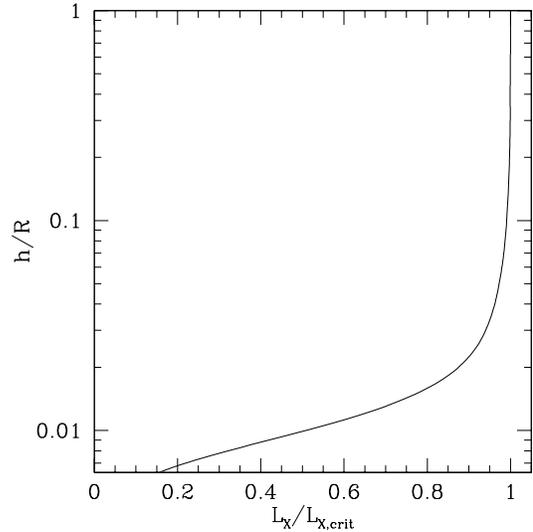}
\caption{Scale height h/R of an X-ray irradiated disk as a function of
  the X-ray luminosity of the central source, for a disk with a
  surface density of $50\,{\rm g\,cm}^{-2}$ around a $10^7$ \Msun\ BH
  at $R=0.2$ pc.  The X-ray luminosity is normalized to $L_{\rm X,\,
    crit}$ from equation (\ref{eq:Lcrit}).  We calculate vertical
  structure models and determine the downscattered X-ray flux
  $F_{X,ds}$ needed to support a disk with a scale height $h/R$.  The
  required X-ray luminosity of the central source is then estimated to
  be $L_X = 2\times 4\pi R^2 F_{X,ds} (R/h)$ (see the text for details).
  Note how a factor of 2-3 increase in $L_X$ increases the
  scale-height of the disk by a factor of $\sim 100$, from $h/R
  \approx 0.01$ to $h/R \approx 1$.}
\label{fig:flux}
\end{figure}

\begin{figure}
\epsscale{1.15} 
\plottwo{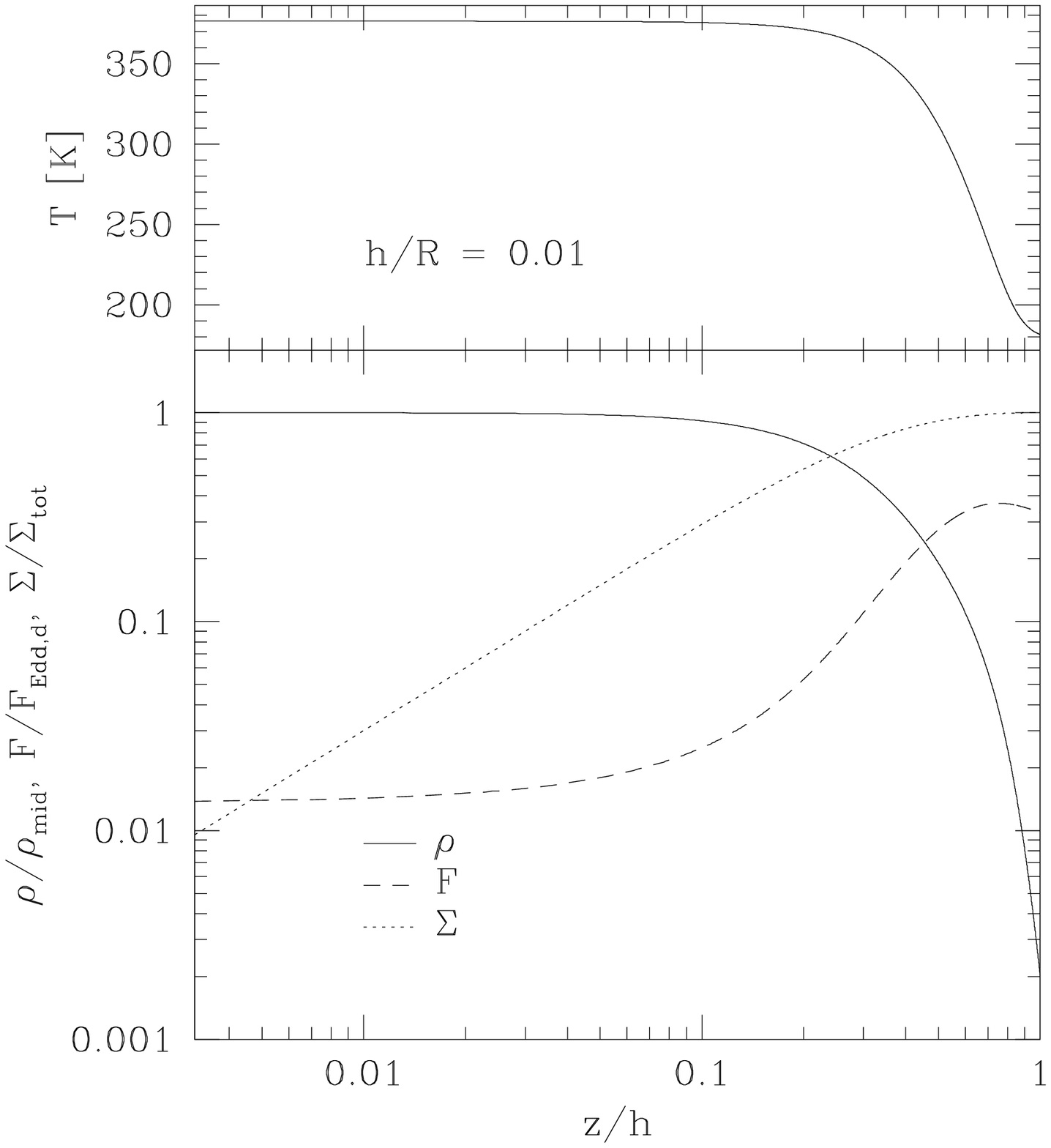}{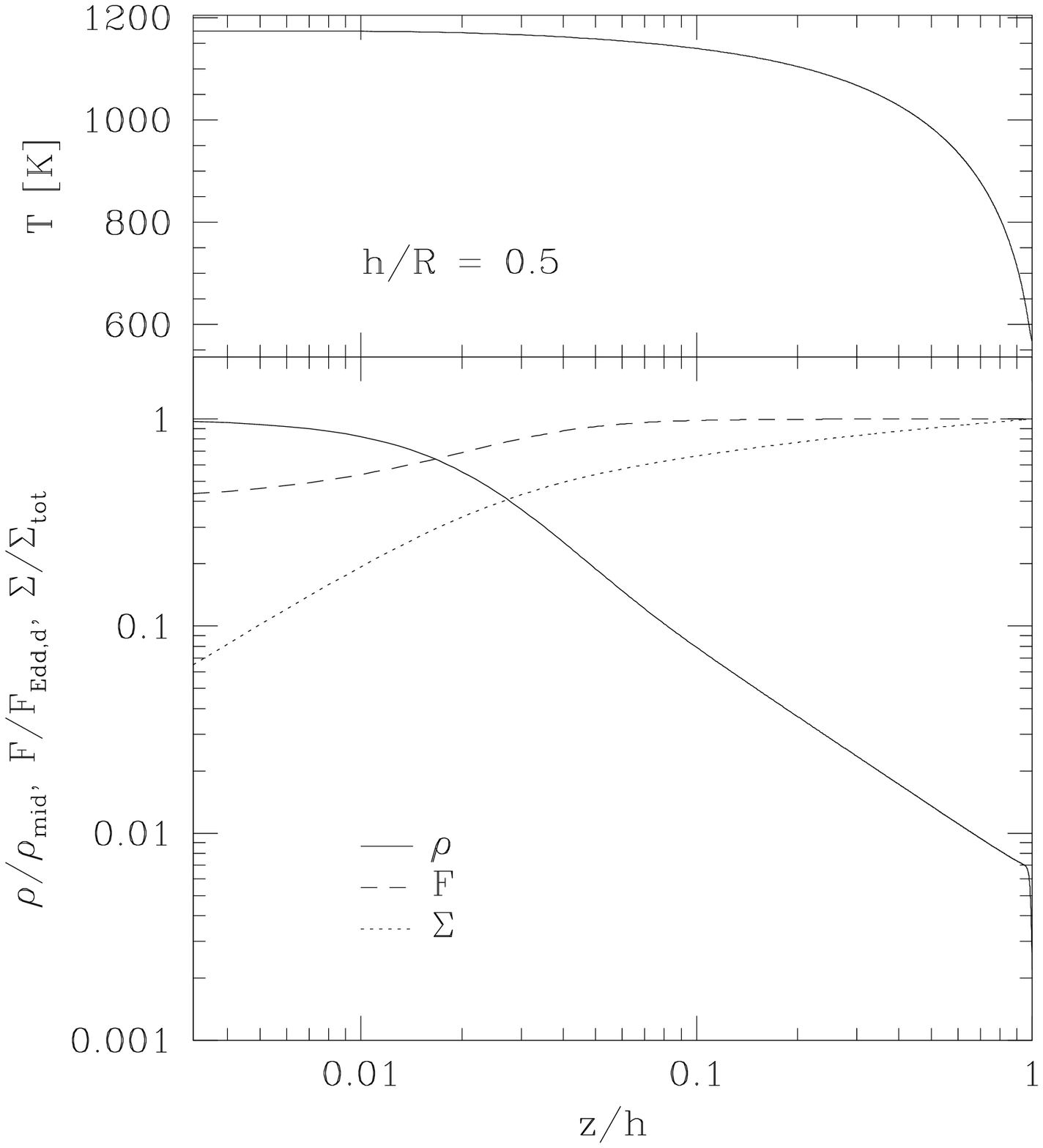}
\caption{Vertical structure of an X-ray irradiated disk around a $10^7
  M_\odot$ BH at R = 0.2 pc.  The incident X-ray flux is adjusted to
  give $h/R = 0.01$ (left) and 0.5 (right). In the upper panel we plot
  the temperature as a function of height.  In the lower panel we plot
  the density (solid line), flux (dashed line) and column (dotted
  line) normalized to the midplane density ($\rho_{\rm mid}$),
  photospheric Eddington flux ($F_{\rm Edd,d} = G M h/(R^3
  \kappa_d)$), and total column $\Sigma_{\rm tot} = 50\,{\rm g\,cm}^{-2}$,
  respectively.  The column, $\Sigma$, is defined to be zero
  at the midplane.  For $h/R = 0.01$, the disk is gas pressure
  dominated. The $h/R=0.5$ solution, on the other hand, is radiation
  pressure dominated ($F/F_{\rm Edd,d} \approx 1$) above $z/h \sim
  0.05$.  The half-mass heights (midplane densities) are 0.002 R ($n
  \approx 1.5 \times 10^{10}\,{\rm cm}^{-3}$) and 0.02 R ($n \approx
  2\times 10^9\,{\rm cm}^{-3}$) for $h/R = 0.01$ and $h/R = 0.5$,
  respectively.  The incident X-ray flux is $F_{\rm X} \approx 3.3
  \times 10^6\,{\rm ergs\,cm^{-2}\,s^{-1}}$ ($L_X/L_{\rm Edd} \approx
  0.01$) for $h/R = 0.01$ and $F_{\rm X} \approx 7 \times 10^6\,{\rm
  ergs\,cm^{-2}\,s^{-1}}$ ($L_X/L_{\rm Edd} \approx 0.03$) for
  $h/R=0.5$.}
\label{fig:0.01-0.5}
\end{figure}

Figure \ref{fig:flux} shows the scale height $h/R$ of an irradiated
disk as a function of the X-ray luminosity of the central source.  The
X-ray luminosity is normalized to $L_{\rm X,\, crit}$ from equation
(\ref{eq:Lcrit}).  Two representative vertical structure models ($h/R
= 0.01$ and $0.5$) are shown in Figure \ref{fig:0.01-0.5}.  In our
calculations, we determine the {\it downscattered} X-ray flux, $F_{X,ds}$, needed to support a
disk with scale height $h/R$.  In Figure \ref{fig:flux}, this is
interpreted as an X-ray luminosity of the central source using $L_X =
2\times 4\pi R^2 F_{X,ds} (R/h)$. The factor of 2 is due to the fraction of
the X-rays downscattered in the atmosphere of the disk and the factor
of $R/h$ is due to the grazing incidence of X-rays (eq. [1]).  The
results in Figures \ref{fig:flux} and \ref{fig:0.01-0.5} are for a
disk with a surface density of $50\,{\rm g\,cm}^{-2}$ around a $10^7$
\Msun\ BH at $R=0.2$ pc. The total column in this case is such that
nearly all of the downscattered X-rays are absorbed (c.f. Figure
\ref{fig:xray_heating}).  The structure of the surface layers of the
disk (e.g., the photospheric height $h$) are independent of the total
column once it exceeds $\approx 20-30$ g cm$^{-2}$. Note that for the
solutions shown in Figures \ref{fig:flux} and \ref{fig:0.01-0.5},
there are no internal sources of heat, such as would be provided by
viscous stresses or star formation; including such internal sources of
heat would lead a minimum $h/R$ even in the absence of irradiation
(unlike in our present solutions in which $h/R \rightarrow 0$ as $L_X
\rightarrow 0$).  This would not significantly change the properties
of our solutions when $L_X \sim L_{\rm X,crit}$.

Figure \ref{fig:flux} shows that for $L_X \ll L_{\rm X,\, crit}$,
the irradiating X-rays have little effect on the structure of the disk,
even in the photospheric layers: the disk remains cold and thin and is
gas pressure dominated throughout.  X-ray irradiation can, however,
still affect the chemistry of the disk in this limit, and can lead to
conditions conducive to water masing ala NGC 4258 (Maloney et
al. 1996).  For low accretion rates, the surface density of an AGN
disk can also be sufficiently small that X-ray irradiation leads to a
transition from molecular to atomic gas (e.g., Neufeld, Maloney \&
Conger 1994).  The gas surface densities in bright AGN are, however,
too large for this to occur.

For $L_X \rightarrow L_{\rm X,\, crit}$, Figure \ref{fig:flux} shows
that the disk puffs up to $h \sim R$.  Even in this limit, however,
most of the mass remains relatively unaffected by the X-ray irradiation
(e.g., the half-mass heights for the models in the left and right
panels of Fig. \ref{fig:0.01-0.5} are 0.002 R and 0.02 R,
respectively).  The transition between X-ray irradiation having
relatively little dynamical effect to forcing the photosphere up to $h
\sim R$ is quite abrupt, with $h/R$ increasing by a factor of $\sim
100$ over only a factor of 2.5 in $L_X$.  Physically, this is because
at the temperatures of interest a gas pressure dominated disk has $h
\ll R$ and only when $L_{X} \sim L_{\rm X,\, crit}$ does radiation
pressure contribute significantly to the pressure support.
Observationally, the rapid change in $h/R$ in Figure \ref{fig:flux}
would correspond to a dramatic change in the nuclear obscuration on
parsec scales for systems with $L_X \gtrsim L_{\rm X,crit}$.

For an irradiating flux of $L_X > L_{\rm x, crit}$, the atmosphere of
the irradiated disk becomes geometrically thick and the outer layers
will be blown away by a combination of the X-ray and UV radiation.
Though a full wind calculation is needed to understand the dynamics of
such a flow in detail, we can roughly estimate the column density
through the unbound material. Since the wind is momentum driven by a
super-Eddington flux, the maximum mass outflow rate is given by
momentum balance, $\dot{M}_{\rm max} v_{\rm esc} = L/c$, where $v_{\rm
esc}$ is the escape velocity from radius $R$ in the disk, which gives
\begin{equation}\label{eq:dotMmax}
  \dot{M}_{\rm max} \sim 2.5 \left( \frac {L}{10^{44}\,{\rm
  ergs\,s}^{-1}}\right) \, R_1^{1/2} \, M_7^{-1/2} \,\Msun\,{\rm
  yr}^{-1},
\end{equation}
where $R_1 = (R/1\,{\rm pc})$ and $M_7 = (M_{\rm BH}/10^7\,\Msun)$.
The obscuring column through such a flow is
\begin{equation}\label{eq:obscuring column}
  N_H \approx \frac {L}{L_{\rm Edd}}m_p^{-1} \kappa_{e}^{-1}
  \approx 10^{24} \frac {L}{L_{\rm Edd}}\ {\rm cm^{-2}}.
\end{equation}
For typical values of $L/L_{\rm Edd}$ in luminous AGN, the column
through the outflowing wind is Compton-thin with $N_H \approx
10^{22}-10^{23}\,{\rm cm}^{-2}$.  Equations (\ref{eq:dotMmax}) and
(\ref{eq:obscuring column}) assume that all of the luminosity of the
central source contributes to driving an outflow.  This is not correct
for the X-ray flux, since the resulting column is Compton thin.
However, once the photosphere of the disk is moderately inflated by
X-ray irradiation, the UV flux will become dynamically important and
will also contribute to unbinding the surface layers of the disk.
Thus the above estimates of the mass loss rate and column are likely
to be reasonable.

\section{Dust Grain Growth in AGN Disks}\label{sec:dust}

As discussed in the previous section, the impact of the X-ray and UV
radiation from a central AGN on its surrounding accretion disk depends
sensitively on the FIR opacity of the disk.  
For temperatures of $\sim 300-10^3$ K, typical models of ISM dust have
opacities of $5-15\,{\rm cm^2\,g^{-1}}$.  We now argue that this
opacity may be enhanced in the near-AGN environment due to 1.
supersolar metallicity and 2. larger dust grains.

If most metals are locked up into dust, enhancements in the
metallicity increase the opacity as $Z/Z_{\odot}$.  From quasar
observations, the broad line region appears to have metallicities that
are a few to ten times solar (Hamann \& Ferland 1999; Baldwin et al.
2003; Dietrich et al.  2003; Nagao, Marconi, \& Maiolino 2006).  Note
that these metallicities probe just the central nuclear regions of
AGN, whose chemical evolution might be significantly different from
the host galaxy (Hamann \& Ferland 1999).  In addition recent work on
the narrow line region of Seyferts and AGN outflows show supersolar
metallicity (Groves, Heckman \& Kauffmann 2006; Arav et. al. 2006).
For Draine's (2003) Rosseland mean opacity of $\kappa_d \approx
15\,{\rm cm^2\,g^{-1}}$, a metallicity enhancement at the level of 2-4
would increase $\kappa_d$ to $\approx$ $30-60\,{\rm cm^2\,g^{-1}}$.
This corresponds to $\kappa_e/\kappa_d \approx 10^{-2}$, in which case
AGN with X-ray luminosities $\gtrsim 0.01-0.03 \, L_{\rm Edd}$ will
have an X-ray flux that is super-Eddington with respect to the dust on
parsec scales.

An independent argument for enhanced FIR opacity in small-scale AGN
disks is that the dust grains may preferentially be larger than in the
local ISM.  
We begin by considering the observational lines of evidence for (and
against) this conclusion.  Recent detections of the 10 \micron\
silicate emission feature in AGN find that the emission feature is
broadened and shifted to slightly longer wavelengths than in the ISM
(Siebenmorgen et. al. 2005; Hao et.  al.  2005; Sturm et al. 2005).
This is analogous to the emission features in Herbig Ae/Be systems
(Bouwman et al 2001) and may indicate different dust properties, in
particular larger or more amorphous grains.  On the other hand, the
silicate absorption features in Type 2 AGN do not appear to require
unusual dust properties, but may require a patchy distribution of dust
(Roche et al.  1991; Hao et al.  2007; Levenson et al. 2007).
Additional contraints on the properties of dust in the vicinity of AGN
can be provided by extinction measurements.  Maiolino et al. (2001ab)
used such measurements to argue that dust grains near AGN may be
significantly larger than interstellar medium grains.  Similarly,
Gaskel et al.  (2004) argue that the reddening law near AGN is
substantially different from any standard reddening law and that the
minimum size for dust grains in the near-AGN environment is $a\approx
0.3\mu{\rm m}$ due to a lack of reddening in the UV.  Using SDSS data
on a large number of AGN, however, Hopkins et al. (2004) argue against
such a Gaskel-like reddening law, and in favor of SMC-like dust.

Although the observational arguments for larger dust grains near AGN
are by no means conclusive, there are theoretical arguments for this
conclusion as well.  To show this, we consider the evolution of dust
grains in AGN disks via accretion of metals from the gas and dust
grain coagulation.  Lacking a detailed model for the ISM on parsec
scales near AGN, we make two simplifying assumptions.  First, we
assume that grain growth is important if it occurs on a timescale less
than the local dynamical time $t_{\rm dyn} \sim \Omega^{-1} \sim 5000
R_1^{3/2} M_7^{-1/2} \, {\rm yrs}$.  The dynamical time is comparable
to the eddy turnover time on the outer-scale, which is $\sim$ the
timescale on which dust would be destroyed by shocks if turbulence is
supersonic in the disk.  Our second simplifying assumption is that we
estimate the gas density in the disk $\rho_g$ in terms of the Toomre Q
parameter for gravitational stability, via $\rho_g =
M_{BH}Q^{-1}/(2\pi R^3) \rightarrow n \approx 6 \times 10^7 Q^{-1} M_7
R_1^{-3}$ cm$^{-3}$.  We assume $Q \sim 1$, as suggested by
theoretical arguments (e.g., Sirko \& Goodman 2003; Thompson et al.
2005) and as is observed in a diverse range of galaxies from local
spirals (e.g., Martin \& Kennicutt 2001) to luminous starbursts (e.g.,
Downes \& Solomon 1998).  Although the ISM in the vicinity of AGN
undoubtedly consists of multiple phases, $Q \sim 1$ should provide a
reasonable guide to the mean ISM densities.  The gas densities implied
by $Q \sim 1$ are also consistent with the densities of $\approx
10^7-10^{10}$ cm$^{-3}$ required to account for the observed water
maser emission from the central parsecs of a number of AGN (see, e.g.,
Lo 2005 for a review).

We begin by considering the timescale for a dust grain to grow
appreciably in mass via accretion of gas-phase metals, which is
\begin{equation}\label{acc}
  t_{\rm acc} \sim \frac {m_{d}}{\rho_m \sigma v_m} \approx \frac 4 3 \frac{a \rho_d}{v_m \rho_m} \approx 100
  \, \frac{a_1 \rho_{d,3} R_1^3Q}{T_3^{1/2} M_7} \, {\rm yrs} \ll
  t_{\rm dyn}
\end{equation}
where $m_{d} = \rho_{d}4\pi a^3/3$ is the mass of the dust grain,
$\sigma = \pi a^2$ is the dust-gas collision cross section, $v_m
\approx 1 \, T_3^{1/2}$ km s$^{-1}$ is the thermal velocity of the
metals at temperature $10^3 T_3$ K (assuming $A \sim 20$), $\rho_d =
3\rho_{d,3}\,{\rm g\,cm}^{-3}$ is the dust grain density, $a = a_1
\micron$ is the radius of the dust grain, and $\rho_m \approx 0.01
\rho_g$ is the mass density in metals.  Note that we take a standard
dust grain density of $3\,{\rm g\,cm}^{-3}$.  Equation (\ref{acc})
shows that dust grains can increase their mass on a timescale
significantly less than the local dynamical time.  Thus even if some
grains (or their mantles) are destroyed, the free metals accrete onto
remaining dust quite rapidly.

The high gas densities in AGN disks also make them a promising
environment for grain coagulation, as is inferred to occur in
protostellar disks (van Boekel et al. 2003).  To show this, we
estimate the characteristic velocity dispersion of dust grains and
show that a significant fraction of small dust grains stick on impact.
The timescale for coagulation is of order the dynamical time so this
process occurs quickly.  

We begin by estimating the velocity dispersion of grains in a
turbulent medium following the argument given by Draine (1985). The
velocity dispersion of grains is determined by the scale on which the
eddy turnover time, $t_{\rm eddy}$, is comparable to the aerodynamic drag
timescale, $t_{\rm drag}$, for grains in gas.  The latter timescale is
given by
\begin{equation}\label{eq:drag}
  t_{\rm drag} \sim \frac {m_{d}}{\rho_g \sigma c_s} \sim
  8 \times 10^6 \rho_{d,3} 
  R_1^3 M_7^{-1} T_{3}^{-1/2}a_{1}Q \,\, {\rm s},
\end{equation}
where $c_s$ is the thermal velocity of the gas.  On the largest
scales, the accretion disk possesses a turbulent velocity approaching
$h/R$ times the orbital velocity, $v_{\rm orb}$.  As in the ISM, these
velocities are likely supersonic compared to the gas sound speed and
so we assume a supersonic cascade with $v_{\rm eddy} \sim (h/R) v_{\rm
  orb} \left(l_{\rm eddy}/h\right)^{1/2}$, where $l_{\rm eddy}$ is the
size scale of the eddy.  The eddy turnover time, $t_{\rm eddy}$, is
given by
\begin{equation}
  t_{\rm eddy} \sim t_{\rm dyn} \left(\frac {l_{\rm eddy}} h\right)^{1/2}.
\end{equation}
As turbulence cascades to smaller and smaller scales, the eddy
velocities decrease.  Eventually, $v_{\rm eddy} \sim c_s$ and
turbulence transitions from supersonic to subsonic.  We define the
eddy turnover time at this scale (when $v_{\rm eddy} = c_s$) as the
sonic time, $t_{\rm sonic}$, which is given by
\begin{equation}
  t_{\rm sonic} \sim t_{\rm dyn} \frac R h \frac{c_s}{v_{\rm
  orb}} \approx 100 T_{3}^{1/2} M_7^{-1} R_1^{2}\left(\frac h R\right)^{-1}\,{\rm yrs}
\end{equation}
The sonic time is still much longer than $t_{\rm drag}$, so that the
dust responds coherently to the eddy.  Below this scale, we assume
that the turbulence is subsonic and incompressible so that it adopts a
Kolmogorov spectrum.  This assumption is conservative in the sense
that continuing the $l^{1/2}$ scaling of supersonic turbulence to
smaller scales would result in smaller grain velocities and thus a
higher probability of dust sticking.  For the Kolmogorov cascade, the
velocities of the turbulent eddies scale like $v_{\rm eddy} \sim c_{s}
(l/l_{\rm sonic})^{1/3}$, where $l_{\rm sonic}$ is the length-scale of
eddies with turnover times of $t_{\rm sonic}$. The velocity of eddies
with $l < l_{\rm sonic}$ or $t < t_{\rm sonic}$ is
\begin{equation}\label{eq:sonic_eddy}
  v_{\rm eddy} \sim c_{s} \left(\frac {l_{\rm eddy}} {l_{\rm sonic}}\right)^{1/3} \sim
c_{s} \left(\frac {t_{\rm eddy}} {t_{\rm sonic}}\right)^{1/2}.
\end{equation}
The most important eddies are those with $t_{\rm eddy} \sim t_{\rm
drag}$.  For $t_{\rm eddy} \ll t_{\rm drag}$, the dust does not
respond to the eddy.  In the opposite extreme, $t_{\rm eddy} \gg
t_{\rm drag}$, all local dust particles will attain a net coherent
motion.  Hence the local random velocity of dust particles is set by
the scale on which $t_{\rm eddy} \sim t_{\rm drag}$, i.e., 
\begin{equation}\label{eq:v_dust}
  v_{d} \sim c_{s} \left(\frac{t_{\rm drag}}{t_{\rm
  sonic}}\right)^{1/2} \sim
  200\,R_1^{1/2}\rho_{d,3}^{1/2}a_{1}^{1/2}\left(\frac h
  R\right)^{1/2}\,Q^{1/2}\,{\rm m\,s}^{-1}.
\end{equation}
Numerical experiments indicate that dust grains that approach one
another at $\lesssim 10$ m/s will stick to each other (Poppe, Blum, \&
Henning 2000), while at much higher velocities $\sim 1 \, {\rm
km\,s}^{-1}$ they are destroyed on impact.

The thickness of the accretion disk in AGN is uncertain.  Models
predict $h/R \sim 0.1-10^{-3}$ on $\sim$ parsec scales (e.g., Thompson
et al. 2005), although this depends on the uncertain viscosity in the
disk.  Taking $h/R \sim 10^{-2}$, we find that the velocity dispersion
for \micron-sized dust grains is $\sim 20$ m s$^{-1}$.  Due to the
$a^{1/2}$ scaling for the random velocity from equation
(\ref{eq:v_dust}), coagulation is more likely for smaller grains,
while fragmentation becomes more likely for larger grains.  In the
previous section we showed that irradiation can extend the photosphere
of the disk up to a height of $\sim R$.  However, most of the mass of
the disk is still contained in a thin layer with $h/R \ll 1$ (see
\S~3).  Thus the above argument about coagulation remains valid for
most of the mass.

We can estimate the timescale for dust coagulation by considering the
fraction of dust grains which approach each other with a low enough
velocity such that they will stick (e.g., Chakrabarti \& McKee 2005).
For this population, the coagulation timescale is
\begin{equation}\label{eq:coag}
  \frac {t_{\rm coag}}{t_{\rm dyn}} \sim 1\, R_1^{3/2} M_7^{-1/2} 
  a_{1} f  
  \left(\frac v {10\,{\rm m\,s}^{-1}}\right)^{-1} Q \, \rho_{d,3}, 
\end{equation}
where $f$ is the fraction of the dust particle population with $v \sim
10\,{\rm m\,s}^{-1}$.  The dust coagulation timescale is roughly the
dynamical time for 1\micron\ grains and much shorter for
submicron-sized grains.  Hence it is likely that a significant
fraction of the dust grains in AGN disks will coagulate to form
larger grains.  As a check on our reasoning, equation (\ref{eq:coag})
can be scaled to conditions appropriate to proto-stellar disks. For $M
\sim 1\Msun$, $R\sim 200$ A.U., and $Q\sim 10$ (Andrews \& William
2007), we find $t_{\rm coag}/t_{\rm dyn} \sim 1$, similar to our
result for AGN disks.  Submm observations which probe dust emission
from the outer parts of protostellar disks indeed find evidence for
significant grain growth, consistent with the inference from equation
(\ref{eq:coag}) (Draine 2006; Andrews \& Williams 2007).

In the ISM of the Milky Way, the timescale for dust to be destroyed by
supernova shocks is significantly shorter than the timescale on which
dust is injected by star formation (e.g., Jones et al. 1994).  Thus
grain growth in the interstellar medium appears crucial for
establishing the observed properties of Galactic dust (e.g., the size
distribution).  We suspect that a similar balance between grain growth
(argued for above) and grain destruction will set the grain size
properties in AGN disks.  In the vicinity of AGN the intense radiation
field, thermal sputtering by hot gas, and shock processing of grains
all likely contribute to grain destruction, and preferentially destroy
smaller dust grains (e.g., Laor \& Draine 1993).  Accurately
estimating the grain destruction time is clearly difficult given the
poor understanding of the physical conditions in parsec-scale AGN
disks.  As a plausible order of magnitude estimate we note that if
the AGN disk is supersonically turbulent on scales where self-gravity
is important (as in the Milky Way), dust will typically encounter a
supersonic shock once every dynamical time.
During such encounters, dust grain are either broken up or have their
outer parts vaporized.  Freed metals quickly recondense
(eq. [\ref{acc}]) and small dust particles coagulate
(eq. [\ref{eq:coag}]).  Equation (\ref{eq:coag}) implies that
coagulation depends on dust grain size, with $t_{\rm coag} \sim t_{\rm
dyn}$ for $a \sim 1 {\rm \mu m}$ on parsec scales.  If the dust
destruction time is indeed $\sim t_{\rm dyn}$, this suggests that the
AGN environment tends to favor the production of \micron-sized dust
grains, although this is clearly uncertain at the factor of a few
level.

If a significant fraction of the dust mass is contained in grains with
sizes of $\sim$ 1\micron\, the FIR opacity will be enhanced,
increasing the importance of the irradiating X-ray and UV flux.  For a
spherical dust grain with radius a, the opacity is maximized when $\pi
a \sim \lambda$, where $a$ is the size of the dust grain and $\lambda$
is the wavelength of interest.  At $T \approx 500$ K, $\lambda \approx
6\micron$ and therefore the dust opacity is maximized for grains with
$a \approx 2 \micron$.  When the opacity is maximized, the cross
section of the dust grain is close to geometric, so that the opacity
is given by
\begin{equation}\label{eq:opac}
  \kappa_{d} = \frac {\pi a^2 f_d}{4/3\pi \rho_{d} a^3} \approx 25 \,
  a_{1}^{-1}\rho_{d,3}^{-1}
  \left(f_d \over 0.01 \right)\,{\rm
    cm^2\,g}^{-1},
\end{equation}
where $f_d \approx 1\%$ (for solar metallicity) is the dust mass to
gas ratio, and as discussed above, $\kappa_d$ increases roughly
linearly with $Z/Z_\odot$.  Note that we have assumed a standard dust
grain density is $\rho = 3\,{\rm g\,cm^{-3}}$.  Large dust grains are
likely full of holes and voids, lowering the density and raising the
opacity.  Equation (\ref{eq:opac}) implies a FIR opacity up to a
factor of few-5 times larger than in the Milky Way (independent of any
enhancement to the opacity due to supersolar metallicity).

\section{Discussion and Conclusions}\label{sec:discussion}

In the presence of a sufficient irradiating X-ray flux, the outer
atmosphere of an initially thin AGN disk puffs up into a geometrically
thick configuration that provides significant obscuration along most
lines of sight.  We estimate that for $L_X \gtrsim L_{\rm X,crit}
\approx 0.01-0.1 \, L_{\rm Edd}$ X-ray heating is dynamically
important for the structure of AGN disks on parsec scales; the upper
end of this range is appropriate for normal ISM dust, while the lower
end of the range is appropriate if, as we have argued, the FIR opacity
in AGN disks is enhanced through a combination of supersolar
metallicities and grain growth (\S 4).  The latter may occur in AGN
disks on parsec scales in direct analogy with massive protostellar
disks.

For $L_X < L_{\rm X,crit}$, irradiation has little dynamical effect on
the properties of AGN disks, while for $L_X > L_{\rm X,crit}$, X-ray
heating likely drives a significant outflow from the surface of the
underlying accretion disk. The transition between these two different
regimes is very abrupt with $h/R$ increasing by a factor of $\sim 100$
with a factor of $\approx 2.5$ increase in X-ray flux
(Fig. \ref{fig:flux}).  Our results thus predict a marked increase in
the presence of small-scale nuclear obscuration for the most X-ray
luminous AGN.  This obscuring material should preferentially lie at
distances of $\sim 0.1-3$ pc from the central AGN because outside
$\sim 3 M_7^{1/2}$ pc the bulge mass dominates the BH mass and so the
irradiating flux is less likely to be locally super-Eddington.  Grain
growth, if it occurs, also occurs preferentially at radii of $\sim
0.1-1$ pc. Using the estimated mass loss rate in a continuum radiation
pressure driven wind, we find characteristic obscuring columns of
$\sim 10^{22-23}\,{\rm cm}^{-2}$ for luminous AGN, in reasonable
agreement with observations (e.g., Risaliti, Maiolino \& Salvati
1999).  Without a more sophisticated model, however, it is
difficult to quantitatively determine the solid angle subtended by
this material, although we expect it to be $\sim \pi$, implying a
significant fraction of Type 2 AGN even at high luminosities.  This is
consistent with, e.g., the implied covering fraction of dusty material
of $\approx 10-50\%$ from the IR fluxes of unobscured Type 1 quasars.

Luminous Seyferts and quasars have X-ray Eddington ratios in the range
required for X-ray irradiation to be dynamically important.  This is
particularly true if the FIR opacity is enhanced in AGN disks.  In a
sample of 35 reverberation mapped AGN, Peterson et al.  (2004) found
that a significant fraction had $L_{\rm Bol} \gtrsim 0.1 L_{\rm Edd}$.
Taking into account the X-ray emission up to $\sim 100$ keV, Seyferts
typically have $L_X/L_{\rm Bol} \sim 0.1 - 0.3$ (e.g., Marconi et
al. 2004), and so we expect quite a few systems will have X-ray fluxes
that are super-Eddington with respect to the dust in a parsec-scale
disk.  At high redshift, bolometric Eddington ratios for luminous AGN
are $\approx 1/3-1$ (e.g., Kollmeier et al. 2006), although quasars
emit a smaller fraction of their flux in the X-ray than do Seyferts
(e.g., Elvis et al. 1994; Marconi et al. 2004) and so their X-ray
Eddington ratios are similar.

We now discuss our model in the context of observations of few
specific systems.  INTEGRAL observations directly probe the high
energy emission from AGN that dominates the Compton heating of
surrounding material.  To focus on a concrete example, the Seyfert 2
galaxy NGC 4388 has an X-ray luminosity of $L_X \approx 4\times
10^{43}\,{\rm ergs\,s}^{-1}$ between $20-100$ keV and is Compton thin
with $N_H \approx 3\times 10^{23}\,{\rm cm}^{-2}$ (Beckman et al.
2006).  NGC 4388 has a central BH mass of $M_{\rm BH} \approx 6 \times
10^6\,\Msun$ (Woo \& Urry 2002), which implies $L_X/L_{\rm Edd}
\approx 0.04$.  This X-ray Eddington ratio and the observed obscuring
column are consistent with our results on the effect of radiation
pressure on parsec-scale disks in AGN.

To consider another concrete example, the prototypical Seyfert 2 NGC
1068 is believed to be radiating at roughly a third of Eddington (Ho
2002; Pounds \& Vaughan 2006).  The inferred X-ray Eddington ratio for
1068 is $L_X/L_{\rm Edd} \approx 0.01$ (e.g., Panessa et al. 2006),
but this is extremely uncertain because 1068 is Compton thick.  For
reasonable intrinsic X-ray Eddington ratios, the X-ray flux of NGC
1068 is quite likely to be dynamically important for the surrounding
material.  Infrared interferometry of NGC 1068 at 8.7\micron\ by Jaffe
et al.  (2004) reveals the presence of a hot optically thick dust
component ($T>800\,{\rm K}$) 0.7 pc from the black hole with a scale
height of $h/R > 0.6$.  The spatial scale and geometric structure of
this obscuring material are consistent with the predictions of our
model.  However, the obscuring column in 1068 exceeds our nominal
prediction of $\sim 10^{22}-10^{23}$ cm$^{-2}$.  In our models, larger
columns would be observed at low inclination when the line of sight
passes through the denser parts of the disks's atmosphere (see
Fig. \ref{fig:0.01-0.5}).

Although radiation pressure is likely to be dynamically significant in
shaping the observed properties of obscuring material in luminous AGN,
it is equally clear that there are a number of systems in which this
is not the case.  In particular, low luminosity AGN such as NGC 4258
(Greenhill et al.  1995) and NGC 3227 (Davies et al. 2006) are far too
sub-Eddington for irradiation to be important.  In such systems, other
modes of obscuration are likely important, perhaps including a warped
disk, obscuration by the host galaxy (e.g. MCG-6-30-15; Ballantyne,
Weingartner, \& Murray 2003), an accretion disk outflow, or inflowing
molecular clouds.  Our prediction is that luminous AGN should
preferentially have a very compact source of nuclear obscuration that
is intrinsically geometrically thick (in contrast to a warped disk).
Continued progress on IR interferometry should allow better
observational probes of the spatial scale of nuclear obscuration in
AGN.  Continuum reverberation mapping between the IR and optical in a
sample of Seyfert 1s already provides evidence that the obscuring
material extends down to the sublimation radius in many systems (e.g.,
Suganuma et al. 2006).

Our calculations have shown that geometrically thin disks on parsec
scales in luminous AGN are likely to be substantially modified by
X-ray irradiation.  Once the surface layers of the disk start to
become geometrically thick, UV heating will become dynamically
important as well.  Since the UV luminosity is larger than the X-ray
luminosity, UV irradiation will ultimately have a significant effect
on the structure of the geometrically thick obscuring material (as in
Pier \& Krolik 1992), although the X-rays will likely, as we have
argued, play the key role in initially thickening this material.  A
proper treatment of the structure of the irradiated parsec-scale disk
thus ultimately demands a self-consistent multi-dimensional
calculation that is beyond the scope of this work (see Krolik 2006 for
axisymmetric hydrostatic calculations of radiation pressure supported
tori).  In light of these uncertainties, the mass outflow rate
(eq. [\ref{eq:dotMmax}]) and obscuring column (eq. [\ref{eq:obscuring
column}]) estimated in \S 3 can only provide an order of magnitude
guide to the mass and column supported by irradiation.  Another
over-simplification in our calculation is that we have modeled the
disk as a uniform medium, which is probably not a good approximation.
In addition to clumping due to self-gravity, radiation supported
atmospheres and winds are subject to instabilities that amplify
inhomogeneities (e.g., Rayleigh-Taylor and photon bubble
instabilities).  This may increase the typical obscuring column at
fixed $\dot M$ relative to that estimated in equation
(\ref{eq:obscuring column}).  Finally, we note that the mass outflow
rate driven by radiation pressure (eq. [\ref{eq:dotMmax}]) may be
substantial relative to the accretion rate onto the central black
hole.  This supports the hypothesis (e.g., Murray, Quataert, \&
Thompson 2005) that feedback from radiation pressure can help regulate
the growth of supermassive black holes in galactic nuclei.

\acknowledgements

We thank the referee, Bruce Draine, for a thorough reading and making
many suggestions which greatly improved this paper.  We thank R.
Antonucci, S. W. Davis, J. Krolik, A. Socrates, and T.  Thompson for
detailed discussions. We thank J. Weingartner for sharing his
numerical calculations on dust opacity.  We thank S. Chakrabarti for
pointing out the importance of grain growth.  We also thank S.
Andrews and J. Williams for providing a preprint of their paper.  P.
C. was supported by the Miller Institute for Basic Research.  E. Q.
was supported in part by NASA grant NNG05GO22H, an Alfred P. Sloan
Fellowship, and the David and Lucile Packard Foundation. N. M. is
supported by NSERC of Canada


\begin{references}

\reference{} 
Andrews, S.~M. \& Williams, J.~P. 2007, submitted to \apj

\reference{}
Antonucci, S. 1993, \araa, 31, 473

\reference{}
Arav, N., Gabel, J.~R., Korista, K.~T., Kaastra, J.~S., Kriss, G.~A.,
Behar, E., Costantini, E., Gaskell, C.~M., Laor, A., Kodituwakku, N.,
Proga, D., Sako, M., Scott, J.~E., \& Steenbrugge, K.~C. 2006, {\it to
  appear in ApJ}, astro-ph/0611928


\reference{}
Baldwin, J.~A., Hamann, F., Korista, K.~T.; Ferland, G.~J., Dietrich,
M., \& Warner, C. 2003, \apj, 583, 649

\reference{}
Ballantyne, D.~R., Weingartner, J.~C., \& Murray, N. 2003, \aap, 409, 503


\reference{}
Beckmann, V., Gehrels, N., Shrader, C.~R., \& Soldi, S. 2006, \apj,
638, 642

\reference{}
Bouwman, J., Meeus, G. de Koter, A., Hony, S., Dominik, C., \& Waters, L.~B.~F.~M.
2001, \aap, 375, 950

\reference{}
Chakrabarti, S. \& McKee, C.~F. 2005, \apj, 631, 792

\reference{}
Davies, R., Thomas, J., Genzel, R., Sanchez, F.~M., Tacconi, L.,
Sternberg, A., Eisenhauer, F., Abuter, R., Saglia, R., \& Bender, R.
2006, {\it accepted by ApJ}, astro-ph/0604125

\reference{}
Dietrich, M., Hamann, F., Shields, J.~C., Constantin, A., Heidt, J.,
Jager, K., Vestergaard, M., \& Wagner, S.~J. 2003,  \apj, 589, 722

\reference{}
Downes, D. \& Solomon, P.~M. 1998, \apj, 507, 615

\reference{}
Draine, B.~T. 1985 in Protostars and Planets II, ed. D.~C. Black \&
M.~S. Matthews(Tucson: Univ.  Arizona Press), 621

\reference{} 
Draine, B.~T. 2003, ARA\&A, 41, 241

\reference{}
Draine, B.~T. 2006, \apj, 636, 1114

\reference{}
Elvis, M., Wilkes, B.~J., McDowell, J.~C.; Green, R.~F., Bechtold, J.,
Willner, S.~P., Oey, M.~S.; Polomski, E., Cutri, R. 1994, \apjs, 95, 1

\reference{}
Gaskell, C.~M, Goosmann, R.~W., Antonucci, R.~R.~J., \& Whysong, D.~H. 2004, \apj, 616, 147

\reference{}
Greenhill, L.~J., Jiang, D.~R., Moran, J.~M., Reid, M.~J., Lo, K.~Y.,
\& Claussen, M.~J. 1995, \apj, 440, 619

\reference{}
Groves, B., Heckman, T., Kauffmann, G. 2006, {\it accepted by MNRAS}, astro-ph/0607311

\reference{}
Hamann, F. \& Ferland, G. 1999, \araa, 37, 487

\reference{}
Hao, L. et al. 2005, \apj, 625, L75

\reference{}
Hao, Lei, Weedman, D.~W., Spoon, H.~W.~W., Marshall, J.~A., Levenson,
N.~A., Elitzur, M., \& Houck, J. R.  2007, \apj, 655, L77

\reference{}
Ho, L.~C. 2002, \apj, 564, 120

\reference{}
Hopkins, P.~F., Strauss, M.~A.; Hall, P.~B.; Richards, G.~T., Cooper, A.~S., Schneider, D.~P., Vanden Berk, D.~E., Jester, S., Brinkmann, J., Szokoly, G.~P. 2004, \aj, 128, 1112

\reference{}
Jaffe, W. et al. 2004, \nat, 429, 47

\reference{} 
Jones, A.~P., Tielens, A.~G.~G.~M., Hollenbach, D.~J., \& McKee, C.~F.\ 1994, \apj, 433, 797 

\reference{}
Kollmeier, J.~A., Onken, C.~A., Kochanek, C.~S., Gould, A., Weinberg,
D.~H., Dietrich, M., Cool, R., Dey, A., Eisenstein, D.~J., Jannuzi,
B.~T., Le Floc'h, E., Stern, D. 2006, \apj, 648, 128


\reference{}
Konigl, A. \& Kartje, J.~F. 1994, \apj, 434, 446

\reference{}
Krolik, J. \& Begelman, M.~C. 1988, \apj, 329, 702

\reference{}
Krolik, J., 2006, submitted to \apj

\reference{} 
Laor, A. \& Draine, B.~T. 1993, \apj, 402, 441

\reference{}
Levenson, N.~A., Sirocky, M.~M., Hao, L., Spoon, H.~W.~W., Marshall,
J.~A., Elitzur, M., \& Houck, J.~R.  2007, \apj, 654, L45

\reference{} 
Lo, K. Y. 2005, ARA\&A, 43, 625



\reference{}
Maiolino, R., Marconi, A., Salvati, M., Risaliti, G., Severgnini, P., Oliva, E., La Franca, F., \& Vanzi, L. 2001a, \aap, 365, 28

\reference{}
Maiolino, R., Marconi, A., Salvati, M., Risaliti, G., Severgnini, P., Oliva, E., La Franca, F., \& Vanzi, L. 2001b, \aap, 365, 37

\reference{}
Maloney, P.~R., Hollenbach, D.~J. \& Tielens, A.~G.~G.~M. 1996, \apj, 466, 561

\reference{}
Marconi, A., Risaliti, G., Gilli, R., Hunt, L.~K., Maiolino, R.,
Salvati, M. 2004, \mnras, 351, 169

\reference{}
Martin, C.~L. \* Kennicutt, R.~C.~Jr. 2001, \apj, 555, 301

\reference{} Murray, N., Quataert, E., \& Thompson, T.~A.\ 2005, \apj, 618, 569 

\reference{}
Nagao, T., Marconi, A., \& Maiolino, R. 2006,  \aap, 447, 157

\reference{}
Neufeld, D.~A.; Maloney, P.~R.; Conger, S. 1994, \apj, 436, L127

\reference{} 
Panessa, F., Bassani, L., Cappi, M., Dadina, M., Barcons, X., Carrera, F.~J., Ho, L.~C., Iwasawa, K. 2006, {\it accepted by A\&A}, astro-ph/0605236

\reference{}
Peterson, B.~M. et al. 2004, \apj, 613, 682

\reference{}
Pier, E.~A. \& Krolik, J.~H. 1992, \apj, 399, L23

\reference{}
Poppe, T., Blum, J., \& Henning, T. 2000, \apj, 533, 472

\reference{}
Pounds, K. \& Vaughan, S. 2006, \mnras, 368, 707 

\reference{}
Ptak, A., Zakamska, N.~L., Strauss, M.~A., Krolik, J.~H., Heckman,
T.~M., Schneider, D.~P., Brinkmann, J. 2006, \apj, 637, 147

\reference{}
Risaliti, G., Maiolino, R., \& Salvati, M. 1999, \apj, 522, 157

\reference{}
Roche, P.~F., Aitken, D.~K., Smith, C.~H., \& Ward, M.~J. 1991,
\mnras, 248, 606

\reference{}
Sanders, D.~B., Phinney, E.~S., Neugebauer, G., Soifer, B.~T., \&
Matthews, K. 1989, \apj, 347, 29

\reference{}
Semenov, D., Henning, T., Helling, C. Ilgner, M., \& Sedlmayr, E.
2003, \aap, 410, 611

\reference{}
Siebenmorgen, R., Haas, M., Krugel, E., \& Schulz, B. 2005, \aap, 436, L5 

\reference{}
Sirko, E. \& Goodman, J. 2003, \mnras, 341, 501

\reference{} 
Suganuma, M., et al. 2006, \apj, 639, 46 
 
\reference{}
Sturm, E. et al 2005, \apj, 629, L21

\reference{}
Thompson, T.A., Quataert, E., \& Murray, N. 2005, \apj, 630, 167



\reference{}
van Boekel, R., Waters, L.~B.~F.~M., Dominik, C., Bouwman, J., de
Koter, A., Dullemond, C.~P., \& Paresce, F. 2003, \aap, 400, L21

\reference{}
Wada, K., \& Norman, C.~A.\ 2002, \apjl, 566, L21 

\reference{}
Woo, J.-H. \& Urry, C.~M. 2002, \apj, 579, 530

\reference{}
Zakamska, N.~L. et al. 2003, \aj, 126, 2125

\reference{}
Zakamska, N.~L. et al. 2005, \aj, 129, 1212

\end{references}
\end{document}